\documentstyle[epsfig]{article}

\newcommand{\ba}{\begin{eqnarray}}
\newcommand{\ea}{\end{eqnarray}}
\begin{document}

\title{Playing Dice with Nuclei: 
Pattern out of Randomness?\footnote{To be published as a feature 
article in Nuclear Physics News, Vol. 11, No. 4 (2001)}} 
\author{Roelof Bijker $^{1}$\footnote{On sabattical leave at Universit\`a 
degli Studi di Genova, Dipartimento di Fisica, Via Dodecaneso 33, 
I-16146 Genova, Italy} and Alejandro Frank $^{1,2 }$ 
\and
$^{1}$ ICN-UNAM, AP 70-543, 04510 M\'{e}xico, DF, M\'{e}xico 
\and
$^{2}$ CCF-UNAM, AP 139-B, 62251 Cuernavaca, Morelos, M\'exico}
\date{September 19, 2001}
\maketitle

\begin{quote}
Many a shaft, at random sent, finds mark the archer little meant! 
\flushright{\em Sir Walter Scott} 
\end{quote}

\begin{quote}
It's chaos, but it's organized chaos.  
\flushright{\em Charles Mingus}
\end{quote}

\section*{Introduction}

In a 1926 letter to Max Born, Albert Einstein wrote: {\em I am convinced 
that God does not play dice}, referring to his dissatisfaction with the 
intrinsic randomness of quantum mechanics. But recent work in nuclear 
structure physics with random interactions, suggests that Nature may 
weave some of its patterns by throwing dice. 

\begin{figure} 
\centerline{\hbox{ 
\epsfig{figure=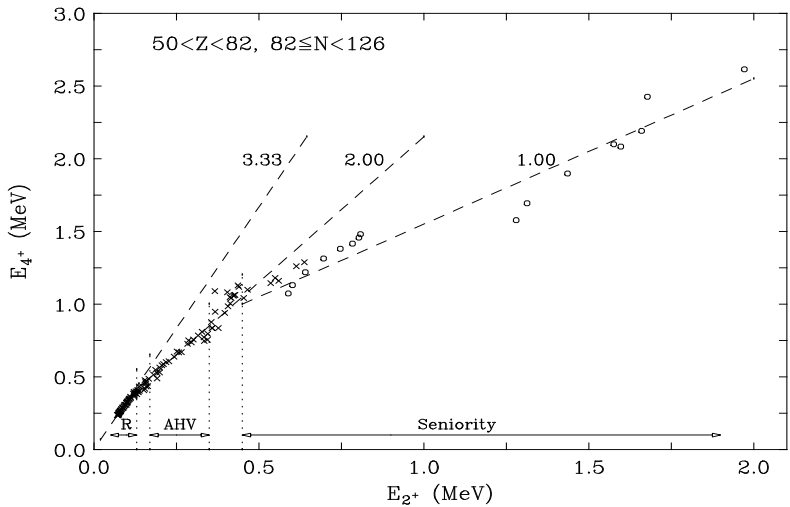,height=0.8\textwidth,width=0.5\textwidth,angle=90} }} 
\vspace{15pt} 
\caption[]{Plot of the excitation energies $E(4_1^+)$ against $E(2_1^+)$, 
showing the tripartite classification into seniority, anharmonic vibrator 
(AHV), and rotor (R) regions \protect\cite{Zamfir}.} 
\label{rick} 
\end{figure}

Low-lying spectra of many-body quantum systems often display a high 
degree of order and regularity. In the case of atomic nuclei, despite 
their complexity and the large number of degrees of freedom involved, 
they often exhibit simple features, such as pairing properties, surface 
vibrations and rotational motion in even-even nuclei. A recent analysis 
of experimental energy systematics of medium and heavy even-even nuclei  
suggests that these nuclei can be classified into just three families, 
namely the pairing, vibrator and rotor regimes \cite{Zamfir}. These areas 
were identified in a plot of the excitation energies of the first-excited 
states $E(4_1^+)$ against $E(2^+_1)$, which show characteristic slopes 
of 1.0, 2.0 and 3.3, respectively (see Figure~\ref{rick}).   

Conventional wisdom is that regularities arise from symmetries of the 
Hamiltonian, which lead to invariances that severely constrain the 
many-body motion.  While some of these symmetries are exact (e.g. 
rotational and reflection invariance), others are approximate 
(e.g. isospin). These global symmetries, however, do not explain by 
themselves the regular patterns observed. Further 
assumptions about the nature of the nucleon-nucleon interaction are 
required. Thus, a strongly attractive pairing force between like 
nucleons has been shown to be responsible for the remarkable 
constancy of the energy of the first excited $2^+$ states in the tin 
isotopes \cite{Talmi}, while deformation and rotational behavior is known  
to arise from an attractive quadrupole-quadrupole interaction between 
neutrons and protons \cite{Talmi}. These striking patterns as well as 
many other correlations have been shown to be robust features of 
low-energy nuclear behavior, which signal the emergence of order and 
collectivity. In every case the patterns arise as a consequence of 
particular forms of the nucleon-nucleon interaction. Thus the general 
belief is that the main features of low-energy 
nuclear spectroscopy and their underlying causes are understood. 

It came as a surprise, therefore, that recent studies of even-even nuclei 
in the nuclear shell model \cite{JBD} and in the interacting boson 
model \cite{BF1} with random interactions displayed a 
high degree of order. Both models showed a marked statistical preference 
($>60 \,\%$) for ground states with angular momentum and parity $J^P=0^+$, 
despite the random nature of the interactions. 

It is the aim of this article to review these new developments in random 
studies of nuclear structure. We first present a brief summary of some 
ensembles of random matrices, and then discuss the main results and their 
possible implications for our understanding of nuclear structure. 

\section*{Random matrix ensembles}

Physicists have long known that generic spectral properties, such as 
average distributions and fluctuations of peaks in neutron-capture 
experiments, or the statistical properties of small metallic particles and 
quantum dots, seem to be independent of the interactions involved, even 
if the data does not precisely match the properties of any one system. 
In order to describe statistical properties of nuclear spectra, Wigner 
developed Random Matrix Theory \cite{Wigner,Porter}, in which the 
Hamiltonian matrix elements are chosen at random, but keeping some global 
symmetries, e.g. the matrix should be hermitean, and be invariant under 
time-reversal, rotations and reflections. 
In his own words: {\em ... the Hamiltonian which governs the 
behavior of a complicated system is a random symmetric matrix, with no 
particular properties except for its symmetric nature}. 
Why is this? In nuclear physics, for example, high energy 
dynamics is assumed to have lost track of its correlations: there is no 
memory of a particular mechanism, or this mechanism turns out to be 
irrelevant in the determination of statistical or generic properties. 
Specifically, the Gaussian Orthogonal Ensemble (GOE) of real-symmetric 
random Hamiltonian matrices closely describes the level repulsion found in 
the distribution of nearest-neighbor spacings of states with the same 
quantum numbers, such as that found in neutron-capture resonances in 
$^{167}$Er, proton resonances in $^{47}$V, and in shell model calculations, 
among many other examples \cite{Brody}. 

It should be noted, however, that the GOE gives rise to a semicircular 
energy level distribution, in contrast to the Gaussian distribution 
found for shell model calculations. An arbitrary choice of the Hamiltonian 
matrix elements, as in GOE, corresponds effectively to random many-body 
interactions in which the maximum order $k$ is equal to the number of 
particles $n$. On the other hand, realistic shell model Hamiltonians 
should contain mostly one- and two-body interactions. A different ensemble 
of random interactions is the so called Two-Body Random Ensemble (TBRE) 
in which the two-body interactions are taken from a distribution of 
random numbers \cite{French,Bohigas}. For $n=2$ particles, TBRE becomes 
identical to GOE, but for $n>2$ particles, the many-body matrix elements 
of the Hamiltonian are correlated and can be expressed in terms of the 
two-body random matrix elements. The expansion coefficients are determined 
by the many-body dynamics of the model space (i.e. angular momentum 
coupling coefficients, coefficients of fractional parentage, etc.). 
The TBRE gives rise to a Gaussian energy level distribution. The transition 
between GOE and TBRE has been studied for the case of seven identical 
nucleons in the $f_{5/2}f_{7/2}$ shell and states with angular momentum 
and parity $J^P=\frac{7}{2}^+$. As the order $k$ of the random 
interactions decreases, the level distribution shows a transition from 
semicircular for $k=7$ (GOE) to Gaussian for $k=2$ (TBRE) \cite{French}. 
For a detailed discussion of the these and other properties of GOE and 
TBRE we refer the reader to the review articles by Brody et al. \cite{Brody} 
and by Guhr et al. \cite{Guhr}. 

\section*{An unexpected result}

Most studies with random matrix ensembles, either GOE or TBRE, involved 
sets of highly excited states with the same quantum numbers, such as 
angular momentum, parity and isospin. In effect, these studies probed 
the eigenvalue distributions near their middle part, and supported the 
general validity of random matrix methods regarding averages and 
fluctuations. Until recently, however, very little attention had been 
paid to correlations among different symmetries and to low-energy behavior, 
i.e. the tails of the distributions. Unlike in the GOE, in the TBRE 
the energy eigenvalues of states with different quantum numbers are 
strongly correlated, since they arise from the same Hamiltonian. 

In a recent development, Johnson, Bertsch and Dean carried out shell 
model calculations 
for even-even nuclei in the $sd$ shell and the $pf$ shell with randomly 
distributed two-body interactions \cite{JBD}. An analysis of the entire 
energy spectrum, i.e. states of all allowed values of the angular momentum 
and other quantum numbers, showed a remarkable statistical preference 
for ground states with $J^P=0^+$, plus the appearance of energy gaps, and 
other indicators of ordered behavior, despite the random nature of the  
two-body matrix elements (both in sign and relative magnitude).  
The unexpected strong dominance of $0^+$ ground states amazed nuclear 
physicists and motivated a large number of investigations. 

These studies explore low-lying features of nuclear structure by addressing 
the following problem. Consider that a certain nucleus is described by 
means of an ensemble of Hamiltonians acting on a given single-particle 
space of valence orbits. The interactions are restricted to two-body, 
satisfy hermiticity, and are invariant under time-reversal, rotations 
and reflections. The two-body matrix elements are taken from a (Gaussian) 
distribution of random numbers with zero mean, so that they are arbitrary 
and equally likely to be attractive or repulsive. 
Next, the many-body Hamiltonian matrices are calculated for each 
value of the angular momentum and diagonalized. The resulting spectrum 
is analyzed for its spectral properties, such as the angular momentum of the 
ground state, the relative position of other yrast states and quadrupole 
transitions among them. This procedure is repeated, let's say 1,000 times. 
What is the result of such a numerical experiment? 

\begin{table} 
\centering
\caption[]{Comparison of the percentage of $J=0$ states in the basis 
and obtained as ground states in TBRE for (i) 6 neutrons in the $sd$ 
shell (the nucleus $^{22}$O) and (ii) for the IBM with $N=16$ bosons.}
\label{lzero} 
\vspace{15pt}
\begin{tabular}{lrr}
\hline
& & \\
& Basis & TBRE \\
& & \\
\hline
& & \\
$^{22}$O &  9.9 $\%$ & 67.7 $\%$ \\
IBM      &  3.3 $\%$ & 63.4 $\%$ \\
& & \\
\hline
\end{tabular}
\end{table}

As an example, here we consider the nucleus $^{22}$O which is described by 
six valence neutrons in the $sd$ shell, which consists of single-particle 
orbitals with $s_{1/2}$, $d_{3/2}$ and $d_{5/2}$ \cite{JBD,BFP1}. 
In Table~\ref{lzero} we show that the percentage of the total number of runs 
for which the ground state has angular momentum $J=0$ is 67.7 $\%$. This 
is a seven-fold enhancement 
with respect to the percentage of $J=0$ states in the model space of 
only 9.9 $\%$. This result is rather insensitive to the model space and 
to the ensemble of two-body interactions \cite{JBD,BFP1,JBDT,DD}. 
In subsequent studies, however, it was shown that the overlap of these 
states with realistic shell model wave functions is small \cite{Horoi}. 

\begin{figure}
\centerline{\hbox{
\epsfig{figure=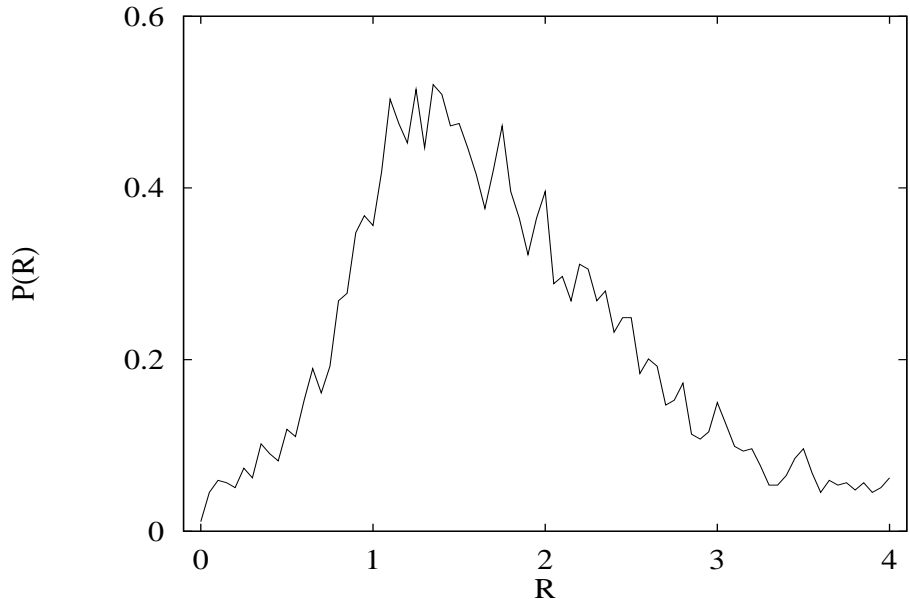,height=0.5\textwidth,width=0.8\textwidth} }}
\vspace{15pt}
\caption[]{Probability distribution $P(R)$ of the energy ratio 
$R=E(4_1^+)/E(2_1^+)$ for six neutrons in the 
$sd$ shell with random two-body interactions.}
\label{sm}
\end{figure}

For the cases with a $J^P=0^+$ ground state, it is of interest to calculate 
the probability distribution of the ratio of excitation energies 
\ba
R &=& E(4_1^+)/E(2_1^+) ~, 
\nonumber
\ea
which constitutes a measure of the fundamental properties of random
nuclei. This energy ratio has characteristic 
values of $1$, $2$ and $10/3$ for the pure seniority, vibrational and 
rotational regions, respectively. Figure~\ref{sm} shows that for six 
neutrons in the $sd$ shell, the probability distribution $P(R)$ has very 
little structure. There is a broad peak between $1 \leq R \leq 2$, with 
a maximum around $1.3$. This suggests that a system of identical nucleons 
on average tends to behave in accord with the seniority regime of 
\cite{Zamfir}. Since the distribution extends to $R=2$, there is some 
evidence, although hardly convincing, for the appearance of vibrational 
structure. On the other hand, for this system there is no occurrence of 
rotational bands \cite{JBD,BFP2}. Recently it was shown, that 
neutron-proton systems do evince such collective behavior if the 
ensemble of two-body interactions is taken from a displaced Gaussian 
distribution (with negative mean) \cite{Zuker}. 

\begin{figure}
\centerline{\hbox{
\epsfig{figure=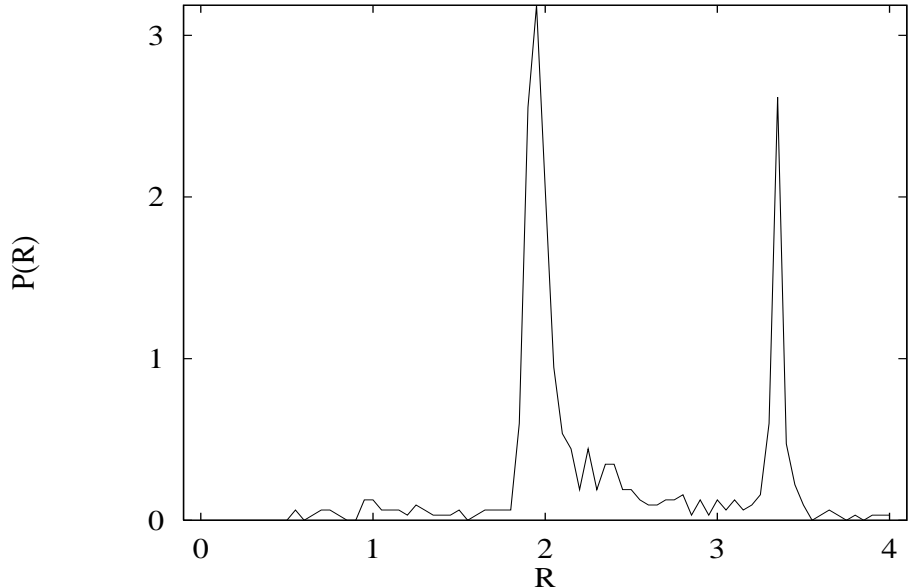,height=0.5\textwidth,width=0.8\textwidth} }}
\vspace{15pt}
\caption[]{Probability distribution $P(R)$ of the energy ratio 
$R=E(4_1^+)/E(2_1^+)$ in the IBM with random 
one- and two-body interactions. The number of bosons is $N=16$.}
\label{ratio12}
\end{figure}

To gauge the robustness of these results, tests were carried out for other 
nuclear models as well. A similar preponderance of $J^P=0^+$ ground states 
was found in an analysis of the Interacting Boson Model (IBM) with random 
interactions \cite{BF1}. In the IBM, collective nuclei are described as a 
system of $N$ interacting monopole and quadrupole bosons \cite{IBM}. 
These bosons reflect the dominant angular momentum components of the 
pairing interaction between identical nucleons. In spite of its simplicity, 
the IBM displays a wide range of collective behavior, which includes shape 
transitional regions. For the case of $N=16$ bosons, a twenty-fold 
enhancement was found, since in 63.4 $\%$ of the cases the ground state 
has $J^P=0^+$, compared to only 3.3 $\%$ of the total number of basis states 
(see Table~\ref{lzero}). Just as for the shell model, the probability 
distribution $P(R)$ of the energy ratio $R$ can be used to look for 
evidence for vibrational and/or rotational structure. 
Fig.~\ref{ratio12} shows a surprising result: the probability 
distribution $P(R)$ displays two very pronounced peaks, one at 
$R \sim 1.9$ and a narrower one at $R \sim 3.3$. 
These values correspond almost exactly to the harmonic vibrator 
and rotor values of 2 and 10/3. A plot of the 
energy ratio and the corresponding ratio of $B(E2)$ values, shows 
a strong correlation between the first peak and the vibrator value  
for the ratio of $B(E2)$ values, as well as between the second peak 
and the rotor value \cite{BF1}. In this case, the evidence for regular 
patterns is not only based on energy systematics, but also on the 
structure of the wave functions. 

The unexpected results found for both the shell model and the IBM, have 
raised fundamental questions in nuclear structure physics and have led 
to a revival of random matrix studies. Naively one might expect most 
properties of these random nuclei to occur stochastically, as conventional 
wisdom suggests that they depend strongly on the particular form of the 
interactions. But are there fundamental properties that remain essentially 
invariant under arbitrary interactions? One of the most striking features 
of these studies is that some of the most hallowed aspects of nuclear 
structure, such as the appearance of $0^+$ ground states, and the 
occurrence of vibrational and rotational motion, can emerge with a strong 
statistical preference from randomly distributed interactions. 

These results, however, were obtained from numerical studies. It is 
necessary to gain a better understanding as to why this happens. What is 
the origin of the regular features which arise from random interactions? 
{\em I am very happy to learn that the computer understands the problem, 
but I would like to understand it too}, as Wigner once quipped. 

\section*{Search for an explanation}

The observed dominance of $J^P=0^+$ ground states was certainly not 
anticipated, considering that there is no obvious pairing character in 
the aleatory forces. The ingredients of these numerical simulations, 
both for the shell model and for the IBM, are the structure of the model 
space, the ensemble of random Hamiltonians, the order of the interactions 
(one- and two-body), and the global symmetries, i.e. time-reversal, 
hermiticity and rotation and reflection symmetry. The latter three 
symmetries of the Hamiltonian cannot be modified, since we are studying 
many-body systems whose eigenstates have real energies and good angular 
momentum and parity. In this section, we discuss the role of some of the 
other ingredients, in search for a possible explanation of this 
phenomenon. 

(i) The dominance of $0^+$ ground states has been shown to be a robust 
property that arises from many different ensembles of random interactions, 
such as the Random Quasi-particle Ensemble which was used in the original 
article \cite{JBD} or the TBRE in subsequent studies \cite{BFP1,JBDT,DD}. 
The inclusion of realistic single-particle energies hardly changes the 
results \cite{JBDT,DD}. Even for ensembles in which the pairing interactions 
have been excluded (set to zero), still more than half of the cases 
have a $0^+$ ground state \cite{JBDT}. 

(ii) Another possibility is that this behavior is related to the 
time-reversal invariance of the random Hamiltonian. 
After all, nuclear pairing, as in electron Cooper pairs in BCS theory, 
involves the filling of time-reversed pairs in doubly degenerate single 
particle states. Since time-reversed states play an important role in 
these favored collective states, it is conceivable that time-reversal 
invariant Hamiltonians induce a built-in preference for $0^+$ ground 
states. To see whether this is indeed the case, it is possible to break 
time-reversal invariance in the random two-body interactions, while still 
maintaining hermiticity and rotational and reflection invariance \cite{Brody}. 
The corresponding ensemble of random two-body matrix elements is the 
well-known Gaussian Unitary Ensemble (GUE). For the case of six identical 
nucleons in the $sd$ shell, the breaking of time-reversal invariance of the  
two-body interactions increases the number of independent random matrix 
elements from 30 to 46. The dominance of $0^+$ ground states turns out to 
increase from 67.7 $\%$ (see Table~\ref{lzero}) to 76.8 $\%$ 
\cite{BFP1}. On the basis of these results, one may conclude that 
time-reversal invariance is not the origin of the dominance of $0^+$ 
ground states. 

(iii) The dependence on the order of the random interactions has been 
investigated in the framework of the IBM \cite{BF2}. It was found that 
whenever the number of bosons $N$ is sufficiently large compared to the 
rank $k$ of the interactions, the spectral properties are characterized by 
a dominance of $0^+$ ground states and the occurrence of both vibrational 
and rotational bands. These band structures appear gradually with increasing 
values of $N/k$. Essentially the same behavior is found for random two- 
and three-body interactions. 

The observed spectral order in systems 
with random interactions cannot be explained by the time-reversal symmetry 
of the interactions, or by the choice of a specific ensemble, nor by the 
order of the many-body interactions. These results suggest that such 
features arise as robust properties of the many-body dynamics of the model 
space and/or of the general statistical properties of random interactions. 
In this respect, we mention two recent developments which may shed some 
light on the problem, and may help to lead toward a possible solution. 

For the case of identical nucleons occupying a single-$j$ shell, an 
explanation has been suggested based on the idea of geometric chaoticity, 
which appears in finite many-fermion systems with complex interactions 
due to exact rotational invariance \cite{MVZ}. It was shown that 
statistical correlations of fermions drive the ground state spin to 
its minimum or its maximum value. The statistical approach predicts 
a smooth behavior of the $0^+$ ground state probability which, for a 
$j^4$ configuration, decreases from $\sim 50$ $\%$ to $\sim 40$ $\%$ with 
increasing values of the angular momentum $j$ of the single-particle orbit. 
However, such a simple statistical approach cannot explain the large 
oscillations observed in the exact numerical results. A recent analysis 
of the structure of the random wave functions lends further support to some 
aspects of this interpretation \cite{Horoi}. 

\begin{figure}
\centerline{\hbox{
\epsfig{figure=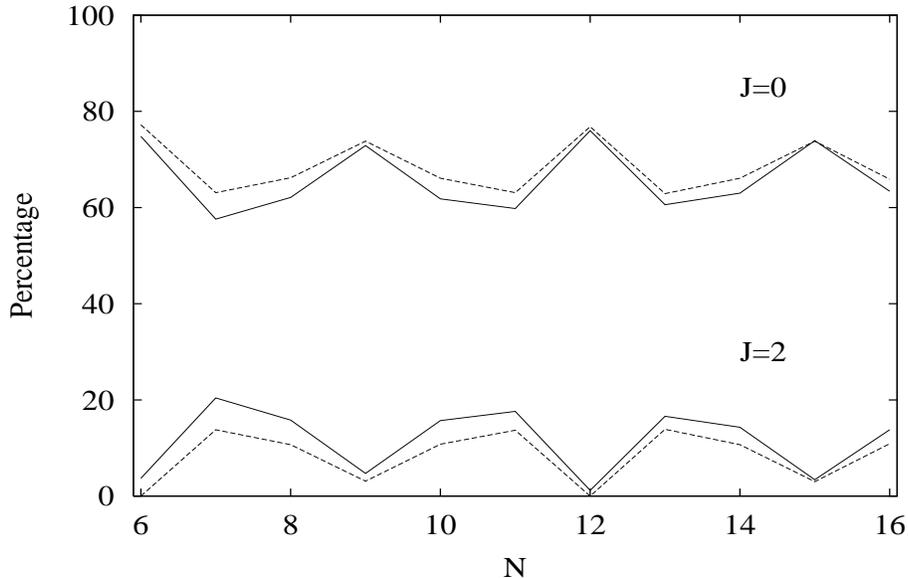,height=0.5\textwidth,width=0.8\textwidth} }}
\caption[]{Percentages of ground states with angular momentum $J=0$ and 
$J=2$ in the IBM with random one- and two-body interactions calculated 
exactly (solid lines) and in mean-field approximation (dashed lines).}
\label{ibmgs}
\end{figure}

In a different development, a Hartree-Bose mean-field analysis of the IBM 
Hamiltonian has been used to associate regions of the parameter space with 
particular intrinsic vibrational states, which in turn correspond to 
definite geometric shapes \cite{BF3}. The results of this analysis indicate 
that there are three basic shapes: a spherical one carried by a single 
state with $J=0$, a deformed shape which corresponds to a rotational band 
with $J=0,2,\ldots,2N$, and a condensate of quadrupole bosons which has a 
more complicated angular momentum content. The ordering of rotational 
energy levels depends on the sign of the corresponding moments of inertia. 
In Fig.~\ref{ibmgs} we show the percentages of ground states with 
$J=0$ and $J=2$ as a function of the total number of bosons $N$. 
A comparison of the results of the mean-field analysis (dashed lines) 
and the exact ones (solid lines) shows good agreement. 
There is a dominance of $J=0$ ground states for $\sim$ 63-77 
$\%$ of the cases. The large oscillations with $N$ are due to the 
contribution of the condensate of quadrupole bosons. The sum of the 
percentages of $J=0$ and $J=2$ ground states hardly depends on the number 
of bosons. The mean-field analysis 
explains both the distribution of ground state angular momenta and 
the occurrence of vibrational and rotational bands. A similar analysis 
has been carried out for the vibron model, an interacting boson model 
to describe the relative motion in 
two-body systems, for which part of the results have been obtained 
analytically \cite{BF4}. 

\section*{Conclusions and outlook}

Numerical simulations for the nuclear shell model and the IBM with 
random interactions suggest that global properties in even-even nuclei, 
such as $J^P=0^+$ ground states (for both models) and the occurrence of 
vibrational and rotational bands (for the IBM) may arise from a much 
broader class of Hamiltonians than the ones usually considered. 
These unexpected results have sparked a large number of investigations 
to explain and further explore the properties of random nuclei. 
Although they have shed light on various aspects of the original problem, 
i.e. the dominance of $0^+$ ground states, in our opinion, no 
definite answer is yet available, and the full implications for nuclear 
structure physics are still to be clarified.  
Open questions include, among others, the properties of higher excited 
states, the transition from ordered, regular 
features to chaos for an ensemble of Hamiltonians, and the emergence 
of collective traits from appropriately constrained random shell-model 
interactions. Other, more general, problems involve the study of the 
statistical properties of randomly interacting many-body systems and 
quantum chaos, such as the spectral properties of random matrix ensembles  
\cite{DD,HAW,Flores}, the structure of wave functions \cite{Kaplan,Kota}, 
the relation with random polynomials \cite{DK}, the matrix elements of the 
Hamiltonian \cite{ZA}, and the connection between statistical and dynamical 
effects in finite many-body systems \cite{MVZ}. 

The study of random matrix ensembles is an exciting interdisciplinary 
field whose universal properties have allowed to establish connections 
between, at first sight, completely unrelated areas of physics and 
mathematics \cite{PT}. It has been known for a long time that random 
matrix ensembles, such as GOE and GUE, reflect universal properties 
of complex systems, ranging from level spacings in slow neutron resonances 
to quantum dots and chaotic billiards \cite{Brody}. Surprisingly, recent 
research suggests a link between the statistical distribution of extreme 
eigenvalues of random matrices and the distribution of prime numbers 
through the Riemann hypothesis \cite{solitary}, thus relating the 
properties of random nuclei discussed in this contribution to 
the most important problem in number theory. 

We conclude with the observation that the study of random phenomena 
in nature is too important to be left to chance. 

\section*{Acknowledgements}

It is a great pleasure to thank Rick Casten, Stu Pittel, David Rowe, 
Piet van Isacker, and Victor Zamfir for stimulating discussions and 
their random suggestions and thoughts. 
This work was supported in part by CONACyT under projects 
32416-E and 32397-E, and by DPAGA-UNAM under project IN106400.

\end{document}